# An RNA foldability metric; implications for the design of rapidly foldable RNA sequences

Wilfred Ndifon[1,]* and Asamoah Nkwanta[2]

[1]Department of Ecology and Evolutionary Biology, Eno Hall 06,

Princeton University, Princeton, NJ 08544

Tel: (609) 258-1678

wndifon@princeton.edu

[2]Department of Mathematics, Carnegie Hall 258,

Morgan State University, Baltimore, MD 21251

Tel: (443) 885-4652

nkwanta@jewel.morgan.edu

30 November, 2005

*To whom correspondence should be sent.

**Abstract**

Evidence is presented suggesting, for the first time, that the protein foldability metric $\sigma = (T_\theta - T_f)/T_\theta$, where $T_\theta$ and $T_f$ are, respectively, the collapse and folding transition temperatures, could be used also to measure the foldability of RNA sequences. The importance of σ is discussed in the context of the <u>in silico</u> design of rapidly foldable RNA sequences.

**1. Introduction**

The temperature dependence of the folding kinetics of selected tRNA sequences is studied, using Monte Carlo (MC) folding simulations and the Monte Carlo Histogram Method (MCHM) [1]. For each tRNA sequence, two characteristic transition temperatures are identified: one at $T_\theta$, when tRNA sequences collapse into compact states, and the other at $T_f$ $(\leq T_\theta)$, when the compact states are converted into the native state. The folding time $\tau_f$ (i.e., the mean first passage time) is also determined for each tRNA sequence, at $T = 37°$ C. $\tau_f$ is found to correlate with $\sigma = (T_\theta - T_f)/T_\theta$, where σ is a metric that describes the kinetic accessibility of the native state [2]. The foldability metric σ was originally introduced by Camacho and Thirumalai [3] in the context of protein folding but, to our knowledge, its applicability had not yet been demonstrated in the context of RNA folding. We now describe in detail the methods and main results of this work and discuss their implications for the design of rapidly foldable RNA sequences.

An RNA folding model described in [4] was used to perform MC folding simulations of selected tRNA sequences (see Table 1), at temperature $T = 37^o$ C. The resulting folding trajectories and the MCHM [1] were used subsequently to compute thermodynamic quantities over a range of temperatures. In particular, the MCHM was used to compute the average overlap <O>, defined as the average Hamming distance from the RNA native structure, and the average compaction <C>, defined as the average number of all RNA base pairs, at temperatures $T = 0^o, 20^o, 40^o, 60^o, 80^o$, and $100^o$ C. The quantities <C> and <O> allowed to estimate the collapse (i.e., $T_\theta$) and folding (i.e., $T_f$) transition temperatures, respectively, as described below.

[Table 1.]

The average compaction was found to have an inverse correlation with temperature, $0^o \leq T \leq 100^o$ C (see Figure 1a). At low temperatures, there is a rapid collapse of the RNA chain into compact states. These compact states are relatively stable and less likely to dissociate, hence the high compaction observed at low temperatures. As the temperature rises, the stability of compact states decreases thereby increasing the likelihood of their dissociation. This explains the low compaction observed at high temperatures. On the other hand, the average overlap was found to correlate positively with temperature, $0° \leq T \leq 100°$ C (see Figure 1b). At high temperatures, there are numerous structures with similar thermal stabilities as the native state. The native state is, therefore, relatively unstable and may be associated with a much smaller basin of attraction on the folding energy landscape. A folding RNA molecule thus spends more time exploring the ensemble of nonnative states, resulting in the high average Hamming distance from the native state. Conversely, at lower temperatures the stability of the native state and, perhaps, the size of its basin of attraction increase, resulting in a corresponding decrease in the average overlap.

[Figure 1]

The folding of proteins is believed to occur in two stages. The first stage involves a rapid collapse of the extended protein chain into an ensemble of compact states [2, 3]. This transition is characterized by the collapse temperature $T_\theta$. It is plausible that RNA sequences exhibit a similar folding behavior [6]. Therefore, assuming two-state kinetics, $T_\theta$ should correspond to the temperature at which $\langle C \rangle = (C_{max})/2$, where $C_{max}$ is the maximum compaction obtained directly from folding trajectories. In the second stage, the ensemble of compact states is explored, followed by a transition from one of these compact states to the native state [2, 3]. This final transition is characterized by the folding temperature $T_f$. Once again, assuming two-state kinetics, $T_f$ should correspond to the temperature at which $\langle O \rangle = (O_{max})/2$, where $O_{max}$ is the maximum overlap obtained directly from folding trajectories. Both $T_f$ and $T_\theta$, in units of Kelvin, were determined for each tRNA sequence and subsequently used to compute $\sigma = (T_\theta - T_f)/T_\theta$. The metric σ was shown previously [3], in the context of protein folding, to correlate with the folding time $\tau_f$. To test the existence of this correlation in the context of RNA folding, $\tau_f$ was determined for all studied tRNA sequences using simulations run at $T$ = 37º C. In Figure 2, $\tau_f$ is plotted against σ. Interestingly, there is a marked positive correlation between $\tau_f$ and σ, suggesting that the protein foldability metric σ may be applicable also to RNA.

[Figure 2.]

RNA molecules play a variety of important functional roles in living cells (e.g., see [7]), mediated by their attainment of specific tertiary structures. The secondary structure is the precursor to tertiary structure - it constitutes both a thermodynamic and a kinetic scaffold on which the tertiary structure is formed [8]. As a result, the thermodynamic and kinetic analysis of RNA secondary structure has been the subject of numerous recent studies (e.g., see [4, 8, 9]). A better understanding of the thermodynamic and kinetic aspects of RNA secondary/tertiary structure will allow the efficient computational design of rapidly foldable RNA sequences having prescribed structural/functional properties [9, 10]. The ensemble of sequences thus designed could be used subsequently as substrates for in vitro selection, thereby decreasing the degree of uncertainty that is inherent in such in vitro selection experiments.

To assist the design of rapidly foldable RNA sequences, computational methods that allow systematic prediction of RNA foldability are highly desirable. A straight forward way to predict the foldability of a given RNA sequence is to determine the sequence's folding time $\tau_f$ via simulation. Unfortunately, the native states of most RNA sequences may not be kinetically accessible within reasonable time scales [6]. Hence, for most RNA sequences, it may not be possible to measure $\tau_f$ via simulation. The metric σ provides an efficient alternative to $\tau_f$, for measuring the foldability of a given RNA sequence since its computability depends only on the length, but not on the foldability, of the sequence under consideration. We briefly describe below how this metric can be applied, in silico, to design RNA sequences that fold rapidly into a desired target secondary structure.

A popular device for designing optimized RNA sequences, in silico, is the flow reactor (e.g., see [11]). To design RNA sequences that fold rapidly into a target structure $S_T$ of size $n$, the following procedure can be performed: (1) A flow reactor is seeded with $N$ randomly generated sequences of length $n$. (2) The minimum free-energy structure $S_i$ and the foldability $\sigma_i$ of each sequence $X_i$ are determined. $S_i$ and $\sigma_i$ are used subsequently to assign to $X_i$ a fitness value $f(X_i)$ using, for example, the following hyperbolic equation:

$$f(X_i) = \frac{1}{0.001 + \sigma_i + d(S_i, S_T)/n}, \tag{1}$$

in which $d(S_i, S_T)$ is the Hamming distance between the parenthesized representations[1] of $S_i$ and $S_T$. A variant of equation (1) has been used previously [12] to investigate the evolutionary dynamics of RNA. Note that the range of theoretical values for both $\sigma_i$ and $d(S_i, S_T)$ is (0, 1), implying that $(2.001)^{-1} \leq f(X_i) \leq 10^3$. (3) Each sequence is replicated, with an error rate equal to the reciprocal of the sequence's fitness. Following replication, *N* of the sequences are selected for continued optimization using, e.g., the stochastic universal sampling method [13]. The sequences that are not selected are allowed to flow out of the reactor. (4) Steps 2 and 3 are repeated for a pre-defined number of times. After a sufficiently large number of rounds of replication and selection, e.g. $10^6$ rounds, the flow reactor would contain an ensemble of optimized sequences that fold rapidly into the target structure, $S_T$. Note that the procedure just described optimizes simultaneously both the thermodynamic and kinetic properties of the designed RNA sequences. Equation (1) can be modified to accord different weights to each of these properties.

---

[1] In the parenthesized representation of an RNA structure, each unbonded base is represented by ".", while for each pair of bonded bases $x$ and $y$, with $x$ closer than $y$ to the 5' end of the RNA, $x$ is represented by "(" and $y$ by ")".

It is worth noting that the evidence establishing a correlation between $\tau_f$ and σ, presented in this paper, is preliminary and not yet conclusive. In particular, both $\tau_f$ and σ should be determined for a significantly larger and more diverse sample of RNA sequences, and their relationship determined. In addition, values for the collapse temperature and the folding temperature computed in this paper are rough approximations of their actual values. More rigorous estimates can be obtained for $T_\theta$, from the temperature dependence of the heat capacity $C_p = (\langle E^2 \rangle_T - \langle E \rangle_T) / T^2$, and for $T_f$, from the slope of $dO/dT$. Note that the plot of $C_p$ versus *T*, over a wide temperature range, will contain one or more peaks corresponding to transitions between various RNA states. In general, $T_\theta$ corresponds to the highest temperature associated with one of these transitions, that is, the transition to the collapsed state. It is hoped that the results reported here will spur more detailed investigations of the relationship between $\tau_f$ and σ, and of potential applications of σ to the design of rapidly foldable, functional RNA molecules.

**Acknowledgements**: This work was funded by NSF-HBCU Grant No. 0236753. We thank Han Liang for suggesting that we incorporate thermodynamic considerations into the method of computing the fitness of an RNA sequence.

**Table and Figure Legends**

Table 1. The tRNA sequences analyzed in this paper. The sequences were retrieved from the tRNA database of Sprinzl et al. [5].

Figure 1. Temperature dependence of the average compaction <C> (a) and average overlap <O>.

Figure 2. Correlation between the folding time $\tau_f$ (i.e., the mean first passage time) and $\sigma = (T_\theta - T_f)/T_\theta$. Note that for each sequence $\tau_f$ was averaged from 100 simulations run at $T=37^\circ$ C.

| Sequence | ID | Organism |
|---|---|---|
| seq1 | *RN1660* | *E. coli* |
| seq2 | *RR1660* | *E. coli* |
| seq3 | *RG1700* | *S. typhi* |
| seq4 | *RH1700* | *S. typhi* |
| seq5 | *RF1580* | *T. thermophila* |
| seq6 | *RH1580* | *T. thermophila* |

Table 1.

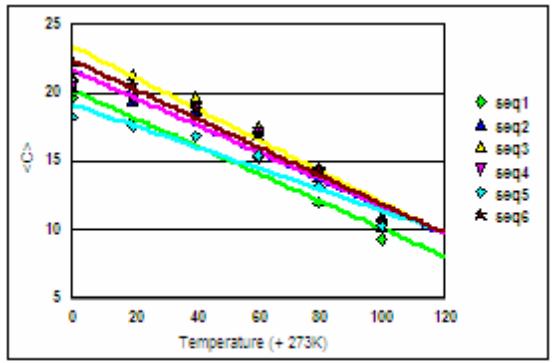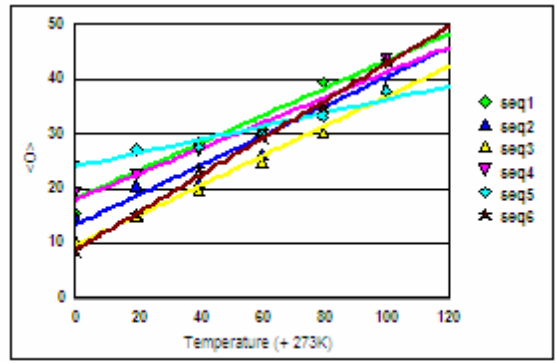

Figure 1.

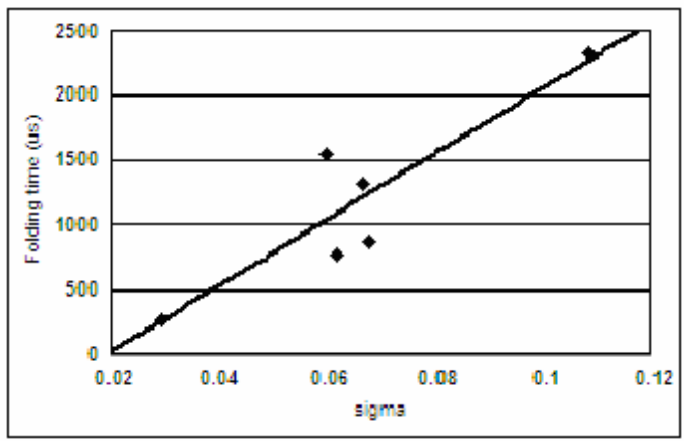

Figure 2.